\documentclass{llncs}

\usepackage[latin1] {inputenc}
\usepackage{amsmath}
\usepackage{amsfonts}
\usepackage{amssymb}

\newtheorem{obs}[lemma]{Observation}

\newcommand{\sD}{\mathcal{D}}
\newcommand{\Prob}[1]{\mathbf{P} \left( #1 \right)}
\newcommand{\Expec}[1]{\mathbf{E} \left[ #1 \right]}

\newcommand{\Probc}[2]{\mathbf{P} \left( #1 \;\left|\; #2 \right.\right)}
\newcommand{\Expf}[1]{\exp\left( #1 \right)}

\begin{document}
\title{\textbf{ MANETS: High mobility can make up for low transmission power }
\thanks{Partially supported by the EU under the EU/IST Project 15964 AEOLUS.}}
\author{
Andrea E.F. Clementi\inst{1}
\and Francesco Pasquale\inst{1}$^\star$
\and Riccardo Silvestri\inst{2}
}

\institute{
Dipartimento di Matematica, Universit\`a di Roma ``Tor Vergata''
\email{$\{$clementi,pasquale$\}$@mat.uniroma2.it}
\and
Dipartimento di Informatica, Universit{\`a}  di Roma ``La Sapienza"
\email{silvestri@di.uniroma1.it}
}
\maketitle

\begin{abstract}
We consider a \emph{Mobile Ad-hoc NETworks (MANET)} formed by $n$ nodes that move independently at random over a finite square region of the plane. Nodes exchange data if they are at distance at most $r$ within each other, where $r>0$ is the \emph{node transmission radius}. The \emph{flooding time} is the number of time steps required to broadcast a message from a \emph{source} node to every node of the network. Flooding time is an important measure of the speed of information spreading in dynamic networks.

We derive a nearly-tight upper bound on the \emph{flooding time} which is a \emph{decreasing} function of the maximal \emph{velocity} of the nodes. It turns out that, when the  node velocity is  ``sufficiently''  high, even if the node transmission radius $r$ is far below the \emph{connectivity threshold}, the flooding time does not asymptotically depend on $r$. This implies that flooding can be very fast even though every \emph{snapshot} (i.e. the static random geometric graph at any fixed time) of the MANET is fully disconnected. Data reach all nodes quickly despite these ones use very low transmission power.

Our result is the first \emph{analytical} evidence of the fact that high, random node mobility strongly speed-up information spreading and, at the same time, let \emph{nodes save energy}.
\end{abstract}

\medskip
\noindent \textbf{Keywords:} Mobile Ad-Hoc Networks, Evolving Graphs, Random Processes, Flooding.

\section{Introduction}
The impact of node mobility in data propagation is currently one of the major issues in Network Theory. The new trend is to consider node mobility as a \emph{resource} for data forwarding rather than a \emph{hurdle}~\cite{KN08,JetAl06}. This is well-captured by the model known as \emph{opportunistic Mobile Ad-Hoc NETworks} (\emph{opportunistic MANET}), an interesting recent evolution of MANET~\cite{PPC06,GT02,JetAl06,ZAZ04}. In opportunistic MANETS, mobile nodes are enabled to communicate even if a route connecting them never exists. Furthermore, nodes are not supposed to have or acquire any knowledge of the network topology (this one being highly-dynamic). Such two features make data communication in opportunistic networks a new challenging research topic from both foundational and practical view-points.

\noindent Inspired by opportunistic MANET, we here consider the \emph{Evolving Graph} yielded by a set of $n$ nodes moving over a finite square. Two nodes can exchange data, at a given time step, if their relative distance at that time step is not larger than a fixed \emph{transmission radius} $r>0$.

\noindent The aim of this work is to investigate the speed of data propagation in the above evolving graph. It is not hard to see that this study must cope with technical problems which are far to be trivial. Just to get an insight of such difficulties, we observe that classic \emph{static} concepts like global \emph{connectivity} and network \emph{diameter} are almost meaningless in this context. On one hand, we can easily construct a sequence of node configurations such that \emph{every} corresponding \emph{snapshot} (i.e. the communication graph at any fixed time step) of the network is not connected while the broadcast task can be performed in a logarithmic number of time steps. On the other hand, it is easy to construct another temporal sequence of node configurations where the snapshot diameter is always 3 while the broadcast task requires $\Theta(n)$ time. Previous  experimental works in this topic show that data communication can benefit from node mobility even though \emph{all} the \emph{snapshots} of the MANETs are not connected~\cite{GT02,PPC06,Z06,KN08,CKNR07}. However, to our knowledge, the only \emph{analytical} evidence of this phenomenon is that proved in~\cite{GT02} concerning \emph{network capacity}, an information-theoretic concept which is not known to be related to the \emph{speed} of data propagation.

\noindent We thus believe that, in order to investigate data propagation on evolving graphs - such as MANETS - a different concept and/or performance measure  should be adopted and investigated.

\noindent In our opinion, a fundamental role in this dynamic world is played by the \emph{Flooding Time}: this is in fact the desired crucial concept/measure. The flooding mechanism is the simple broadcast communication protocol where every \emph{informed} node sends the source message at every time step (a node is said to be \emph{informed} if it knows the source message). The flooding time of an evolving graph is the first time step in which all nodes are informed. The flooding time is a natural lower bound for any broadcast protocol and it represents the maximal speed of data propagation: the same role of the diameter in static networks. Flooding time of some classes of \emph{Markovian Evolving Graphs}~\cite{AKL08} has been recently studied in~\cite{CMMPS08,CMPS09}.

\noindent Our work provides an analytical study of the flooding time of a natural and relevant class of MANETS: we prove a nearly-tight bound on the flooding time showing that high, random node mobility can \emph{dramatically speed-up} data propagation with respect to the corresponding static model. This can be seen as a strong and somewhat final improvement of our previous work~\cite{CMPS09} where we (only) proved that  (low) random node mobility does not \emph{slow-down} data propagation.

\medskip
\noindent \textbf{Our MANET model: An informal definition.}  We consider a model of mobile networks called \emph{geometric Markovian Evolving Graphs}, i.e., \emph{geometric-MEG}~\cite{CMPS09}. It is the discrete version of the well-known \emph{random-walk} model~\cite{CBD02,DMP08,G87} and  can also be viewed as the \emph{walkers} model~\cite{DMP08} on the square.

\noindent In \emph{geometric-MEG}, nodes (i.e. radio stations) move over a finite square region of the plane and each node performs, independently from the others, a sort of \emph{Brownian} motion. In our model we make time and space discrete (see Section~\ref{sec::themodel} for details). The mobility parameter is the \emph{move radius} $\rho$. At every time step, a node moves uniformly at random to any point which is within distance $\rho$ from its current position; so $\rho$ determines the \emph{maximal node velocity}. At any time there is an edge (i.e. a bidirectional link) between two nodes if they are at distance at most $r$.

\noindent It turns out that a Geometric-MEG is a temporal sequence of random disk graphs (\emph{random geometric graphs}~\cite{Penrose}). When $r$ is over the \emph{connectivity threshold}, such random graphs are with high probability\footnote{As usual, we say that an event $\mathcal{E}$ occurs \emph{with high probability} if $\Prob{\mathcal E} \geqslant 1 - 1/n^{\Theta{(1)}}$.} (in short, \emph{w.h.p.}) connected and have diameter $D < n$ that depends on $n$ and $r$. In our previous work~\cite{CMPS09}, we proved that when $\rho < r$ (so we are in the case of \emph{low node mobility}) and the latter is over the connectivity threshold, then the flooding time of geometric-MEG is w.h.p. asymptotically equivalent\footnote{Actually, our previous bound leaves an $O(\log\log n)$ gap in a small range of the network parameters.} to the diameter of the corresponding snapshots. Under the assumption $\rho <  r$, it is not hard to show~\cite{CMPS09} this is \emph{the best the flooding can achieve}: the result is in fact asymptotically tight. It thus follows that, in the \emph{slow} case, random node mobility does not significantly affect the flooding time with respect to the static case.

\medskip
\noindent \textbf{Our results.} In this work, we consider the case $\rho > r$. This can viewed as a model for opportunistic MANETS where message transmission is very difficult due to critic environment conditions and/or poor node transmission power while node velocity is high and random. We are thus motivated by the futuristic scenario of mobile wireless sensor environments composed of myriads of tiny nodes dispersed in the environment and subject to high, unpredictable mobility as a consequence of environmental dynamics such as wind, storms or water streams. To our knowledge, the impact of such a high node mobility on the speed of data propagation has never been considered, at least from a foundational perspective.

\noindent We provide a nearly-tight bound on the flooding time for such a case. Let $G$ be a Geometric-MEG of $n$ nodes over a square of edge size\footnote{For clarity's sake, we choose here to keep node density constant as the number of nodes grows.} $\sqrt n$, transmission radius $r$ and move radius $\rho$ such that $r \geq r_0$ and $\rho \ge c\sqrt{\log n}$, where $r_0$ and $c$ are sufficiently large positive constants. Then, w.h.p., flooding in $G$ is completed within

\[
\mathcal{O}\left( \frac{\sqrt n}{\rho} + \log n \right)  \ \ \mbox{ time steps.}
\]

\noindent It is not hard to show that, for $\rho \geqslant r$, the expected flooding time is $\Omega( \sqrt n / \rho )$~\cite{CMPS09}, so our upper bound is nearly-tight and becomes tight whenever the flooding time is $\Omega(\log n)$.

\noindent When the transmission radius $r$ is over the connectivity threshold (i.e. $\Theta(\sqrt{\log n})$), our bound implies that if the move radius $\rho$ is (asymptotically) higher than $r$, the flooding time is (asymptotically) smaller than the diameter of the snapshots. When $r$ is smaller than the connectivity threshold (say, it is a constant $r_0$) while $\rho \geqslant c \sqrt{\log n}$, then our bound implies that flooding can be efficiently completed despite the snapshot of every time step is formed by several connected components of small sizes~\cite{Penrose,GK99}.

\noindent In general, our upper bound says that, in this case, \emph{the flooding time does not asymptotically depends on the transmission radius}. This fact has important technological consequences in the futuristic scenario of large, high-mobile MANETS. The two major goals in MANETS are: i) guarantee good and fast data communication, and ii) minimize node energy-consumption (which is clearly an increasing function of $r$). It is well-known  that in classic, (static or low-mobile) MANETS such two goals are in contrast with each other and, thus, a suitable trade-off must be determined (actually, optimizing this trade-off is currently a major research issue in ad-hoc networking~\cite{A05,KKKP00,SB03}). In particular, we know that~\cite{Penrose,GK99}, in order to guarantee global connectivity (and thus data communication) in static random geometric graphs, the transmission radius must be $\Omega(\sqrt{\log n})$, so it must \emph{increase  with the network size}. 

\noindent In this context, our bound is a strong mathematical evidence of the fact that, when node mobility is relatively high and random, the two above goals are not competing anymore. We can achieve fast data-forwarding by using small transmission radius (so, saving node energy). More importantly, the transmission radius can be an \emph{absolute constant} and, so, it does not need to increase as the network size does. The technology of node transmitter devices can be thus \emph{scalable}. Observe that node mobility in such opportunistic networks is due to the \emph{host} mobility which is often fully independent from sensor devices: high sensor mobility does not (necessarily) imply high energy consumption~\cite{JetAl06,PPC06}.

\medskip
\noindent \textbf{Adopted Techniques.} The bound in~\cite{CMPS09} on the flooding time for the \emph{slow} case is achieved thanks to the expanding properties of the \emph{connected} snapshots of the geometric-MEG which are, in turn, guaranteed by two facts: 1) The stationary node distribution at every time step is almost uniform; and 2) the transmission radius $r$ is over the connectivity threshold. In particular, they imply that, starting from the second time step, the number of informed nodes is large enough to apply standard Chernoff-like bounds. This allowed us to evaluate the number of \emph{new} informed nodes at any successive step. The role of node mobility is thus shown to have a negligible impact on the flooding process.

\noindent This scenario is no longer true when $r$ is below the connectivity threshold (say constant). The snapshots of the geometric MEG are very sparse and disconnected and, hence, their expanding properties are very scarce. This results into a relatively-long initial phase (called \emph{Bootstrap}) of the flooding process where the number of informed nodes is not large enough to get useful concentration results from Chernoff-like bounds. 

\noindent When $\rho >>r$, the flooding process is mainly due to \emph{node mobility} that, roughly speaking, brings the source information \emph{outside} the small connected components of the sparse snapshots: We provide a clean analytical description of this phenomenon. A key-ingredient here is a probabilistic analysis of the Bootstrap. We present a set of probabilistic lemmas for \emph{almost-increasing random processes} that allows us to evaluate, at every time step, the number of new informed nodes even when the latter has a small expected value. The
rather general form of such lemmas might result useful in other similar situations where Chernoff's-like bounds are useless.

\section{The Node Mobility Model} \label{sec::themodel}
We consider  a model of dynamic graphs, introduced in~\cite{CMPS09}, that is a discrete version of the \emph{random walk mobility} model for radio networks~\cite{CBD02}. In the latter model, nodes (i.e. radio stations) move on a bounded region of the plane (typically a square region) and each node performs, independently from the others, a sort of Brownian motion. At any time there is an edge (i.e. a bidirectional connection link) between two nodes if they are at distance at most $r$ ($r$ represents the \emph{transmission radius}). In our model we discretize time and space. We choose to keep the density constant (i.e. the ratio between the number of nodes and the area) as the number $n$ of nodes grows. The region in which nodes move is a square of side $\sqrt{n}$ and the density equals to $1$. We remark that this choice is only for clarity's sake, and all our results can be scaled to any density $\delta(n)$. The nodes can assume positions whose coordinates are integer multiple of a resolution coefficient $\epsilon > 0$. Formally, nodes move on the following set of points
$$
L_{n,\epsilon} = \left\{ (i\epsilon, j\epsilon) \;|\; i,j \in \mathbb{N} \wedge i, j \leqslant \frac{\sqrt{n}}{\epsilon} \right\}
$$
At any time step, a node can move to one of the positions of $L_{n,\epsilon}$ within distance $\rho$ from the previous position. The positive real number $\rho$ is a fixed parameter that we call \emph{move radius}. It can be interpreted as the maximum velocity of a node\footnote{Indeed, a node can run through a distance of at most $\rho$ in a unit of time.}. Formally, we introduce the \emph{move graph} $M_{n, \rho, \epsilon} = (L_{n,\epsilon}, E_{n,\rho,\epsilon})$, where
$$
E_{n,\rho, \epsilon} = \{ (\mathbf{x}, \mathbf{y}) \;|\; \mathbf{x}, \mathbf{y}\in L_{n,\epsilon}\;\; d(\mathbf{x}, \mathbf{y}) \leqslant \rho \}
$$
and $d(\cdot, \cdot)$ is the Euclidean distance. A node in position $\mathbf{x}$, in one time step, can move in any position in $\Gamma(\mathbf{x})$, where $\Gamma(\mathbf{x}) = \{ \mathbf{y}\; |\; (\mathbf{x}, \mathbf{y}) \in E_{n,\rho,\epsilon}\}$. The nodes are identified by the first $n$ positive integers $[n]$. The time-evolution of the movement of a single node $i$ is represented by a Markov chain $\{P_{i,t}\; ;\; t\in\mathbb{N}\}$ where $P_{i,t}$ are random variables (in short \emph{r.v.}) whose state-space is $L_{n,\epsilon}$ and
$$
\Prob{P_{i,t+1} = \mathbf{x} } \quad = \quad
\left\{
\begin{array}{ll}
\frac{1}{|\Gamma(P_{i, t})|} & \mbox{if $\mathbf{x} \in \Gamma(P_{i, t})$}\\
0 & \mbox{otherwise}
\end{array}
\right.
$$
In other words, $P_{i,t}$ is the position of node $i$ at time $t$. Thus, the time evolution of the movements of all the nodes is represented by a Markov chain $\mathcal{P}(n, \rho, \epsilon) = \{P_t \: :\; t\in \mathbb{N} \}$ whose state space is $L_{n,\epsilon}\times L_{n,\epsilon}\times \cdots \times L_{n,\epsilon}$ ($n$ times) and
$$
P_t  = (P_{1,t}, P_{2,t},\ldots, P_{n, t})
$$
Let us fix a \emph{transmission radius} $r > 0$. A \emph{geometric-MEG} is a sequence of random variables $\mathcal{G}(n, \rho,r, \epsilon) = \{ G_t \;:\; t \in \mathbb{N}\}$ such that $G_t = ([n], E_t)$ with
$$
E_t = \{ (i, j) \;|\; d(P_{i, t}, P_{j, t}) \leqslant r\}
$$

\noindent As for the stationary case, we observe that the stationary distribution $\pi_i$ of Markov chain $\{P_{i,t}\; ;\; t\in\mathbb{N}\}$ is (see~\cite{AF99})
$$
\pi_i(\mathbf{x}) \quad =\quad \frac{|\Gamma(\mathbf{x})|}{\sum_{\mathbf{y}\in L_{n,\epsilon}} |\Gamma(\mathbf{y})|}
$$
Moreover, the stationary distribution of $\mathcal{P}(n, \rho, \epsilon)$ is the product of the independent distributions $\pi_i$ for all $i\in [n]$. We say that a geometric-MEG $\mathcal{G}(n, \rho, r, \epsilon) = \{ G_t \;:\; t \in \mathbb{N}\}$ is a \emph{stationary} geometric-MEG if the underlying $P_0$ is random with the stationary distribution of the Markov chain $\mathcal{P}(n, \rho, \epsilon) = \{P_t \: :\; t\in \mathbb{N} \}$. Notice that if $\mathcal{G}(n, \rho, r, \epsilon) = \{ G_t \;:\; t \in \mathbb{N}\}$ is a stationary geometric-MEG then all r.v. $G_t$ are random with the same probability distribution that we call \emph{stationary distribution} of $\mathcal{G}(n, \rho,r, \epsilon)$.

\noindent In the rest of the paper, we will always assume that the move radius $\rho$ is not larger than $\sqrt{n}$.

\section{Bounding the Flooding Time}
In the flooding mechanism, every node that is informed sends the source message at every time step: so, all nodes that are within distance $r$ from an informed node will be informed at the next time step. For the sake of simplicity, every time step is divided into two consecutive actions: i) the \emph{move action}, where nodes make their random move, and ii) the \emph{transmission action}, where the informed nodes send the source message. Clearly this assumption does not affect the asymptotical bound on the flooding time.

\noindent Our result can be formally stated as follows.

\begin{theorem} \label{thm:geoflooding}
Let $\mathcal{G}(n, \rho, r, \epsilon)$ be a stationary geometric-MEG. If $r \geqslant r_0$ and $\rho \geqslant c \sqrt{\log n}$ for sufficiently large constants $r_0$ and $c$, then the flooding time is w.h.p.
$$
\mathcal{O} \left( \frac{\sqrt{n}}{\rho} + \log n \right)
$$
\end{theorem}

\subsection{Proof's Overiew}
The proof consists of a probabilistic analysis of the number of new informed nodes at every time step of the flooding process. In order to cope with this analysis, the temporal process is organized in three consecutive \emph{phases}. Even though it is likely that in the real process these phases happen simultaneously rather than consecutively, our analysis yields the desired upper bound. The phases depend on the current number of informed nodes and on the ``locality-degree'' of the process. As for the latter, we need to partition the square into equal \emph{supercells}, i.e., subsquares of side length $L = \Theta(\rho^2)$. This partition guarantees that any node $v$ in a supercell $S$, after the move-action, can reach \emph{any} position in \emph{any} neighboring supercell with \emph{almost-uniform} probability. Another crucial property yielded by the partition is that, for the first - say - $\mathcal{O}(n)$ time steps, every supercell will contain $\Theta(\rho^2)$ nodes, w.h.p.

\smallskip
\noindent \textbf{The Bootstrap Phase.} In this initial phase, we start our analysis focussing on what happens inside the neighborhood of the supercell $S_0$ containing the source, i.e., the supercell set $N(S_0)$ formed by $S_0$ and its adjacent supercells. We can say that, with positive-constant probability, $S_0$ contains $\Theta(r^2)$ informed nodes after the first time step. Observe that this is the crucial analysis point where we need to go from \emph{positive-constant probability} to \emph{high probability} and we cannot use Chernoff-like bounds. Indeed, in the successive time steps $t > 0$ of this phase, we consider the flooding-rate inside the supercell $S'_t$ having the maximal number of informed nodes at time step $t$. We will then prove that, after $t = \mathcal{O}(\log n)$ time steps, w.h.p., there will be (at least) one supercell \emph{quasi-informed}, i.e., it will have $\Theta(\rho^2)$ informed nodes.

\smallskip
\noindent \textbf{The Spreading Phase.} After the Bootstrap, we can thus assume (w.h.p.) that there is (at least) one supercell quasi-informed. We can thus look at the flooding from a quasi-informed supercell to its adjacent ones. We show that, w.h.p., if a supercell is quasi-informed at a given time step, then all its adjacent supercells will be quasi-informed within the next time step. Since we prove that the \emph{boundary} of any supercell set $D$ has size at least $\Omega(\sqrt{|D|})$, it turns out that this flooding phase makes all the supercells quasi-informed within $\mathcal{O}(\sqrt{n}/\rho)$ time steps.

\smallskip
\noindent \textbf{The Filling Phase.} At the end of the previous phase, we thus have w.h.p. all supercells quasi-informed. The Filling phase consists of the sequence of time steps required to get all supercells informed. We prove that, w.h.p., this final process can be completed in $\mathcal{O}(\log n)$ time steps.

\subsection{Preliminaries}
We need to introduce the following notions.

\begin{itemize}
\item The square is partitioned into squared \emph{supercells} of side length $L$ with
$$
\frac{\rho}{3 \sqrt{2}} \leqslant L \leqslant \frac{\rho}{2 \sqrt{2}}
$$
\item Every supercell is partitioned into squared \emph{cells} of side length $\ell$ with
$$
\frac{r}{1 + \sqrt{2}} \leqslant   \ell \leqslant \frac{r}{\sqrt{2}}
$$

\item The neighborhood $N(S)$ of a supercell $S$ is the set of supercells formed by $S$ and all its adjacent supercells.
\smallskip
\item A supercell is \emph{quasi-informed} at time $t$ if it contains $\gamma \rho ^2$ informed nodes at that time, where $\gamma$ is a suitable positive constant that will be specified later.
\end{itemize}

\noindent We say that the \emph{density condition} holds at time $t$ if, for every supercell $S$, the number of nodes in $S$ at time $t$ is at least $\eta \rho^2$, for a suitable constant $0< \eta < 1$.  Let $\mathcal{D}$ be the following event: the density condition holds for every time step $t = 0, 1,\ldots, n$.

\noindent The proof of the following lemma is omitted since it is an easy consequence of the almost uniformity of the stationary distribution of geometric-MEG.

\begin{lemma} \label{lm::density}
Let $\mathcal{G}(n, \rho, r, \epsilon)$ be a stationary geometric-MEG. If $\rho \geqslant c\sqrt{\log n}$ for a sufficiently large constant $c$, then the probability of event $\mathcal{D}$ is at least $1-1/n^4$.
\end{lemma}

\noindent In the  rest of this section, we will tacitely assume that event $\mathcal{D}$ holds. Thanks to the previous lemma, since we are conditioning w.r.t. an event that holds w.h.p., the corresponding unconditional probabilities are affected by only a negligible factor.

\noindent For the sake of simplicity, we will use the following probability notations. For an event $\mathcal{E}$ and a r.v.
$X$,  the notation
\[
\Probc{\mathcal{E}}{X} \ \leqslant \ p
\]
means that, for every possible value $x$ of $X$, it holds
\[
\Probc{\mathcal{E}}{X=x} \ \leqslant \ p
\]

\subsection{The bootstrap}
We now provide an upper bound on the time required to get at least one supercell quasi-informed. We will prove the bound for $r = r_0$ where $r_0$ is a sufficiently large constant. Observe that, since the flooding time is a non-increasing function of the transmission radius $r$, the same upper bound holds for any $r \geqslant r_0$ as well.

\noindent The following lemma will be used to evaluate the number of new informed nodes after an initial sequence of consecutive time steps. Notice that the r.v. $X_t$s can be mutually dependent.

\begin{lemma}[Almost-increasing random processes]\label{lemma:almostincreasing}
Let $\{ X_t \,:\, t \in \mathbb{N} \}$ be a sequence of random variables with $X_0 = 1$. Assume that two real values $\alpha > 1,\, 0 < \beta < 1$, a positive integer $M$, and a probability $p \in (0,1)$ exist such that for every $t \in \mathbb{N}$ it holds that
\begin{eqnarray}\label{eq:hyp}
\Prob{X_{t+1} < \alpha \, X_t \;|\; X_t < M,\, X_{t-1} , \dots, X_1 } & \leqslant & p \\[2mm]
\Prob{X_{t+1} \geqslant \beta \, X_t \;|\; X_t , \, X_{t-1} , \dots, X_1 } & = & 1
\end{eqnarray}
If $p < \frac{\log \alpha}{e \, \log(\alpha / \beta)}$ then for any $t \geqslant \frac{\log M}{\log \alpha - e \, p \log(\alpha / \beta)}$ it holds that
$$
\Prob{ \bigcap_{i=1}^t \{X_i < M\}} \leqslant \Expf{-pt}
$$
\end{lemma}
\proof For every $t$ define the binary r.v. $Y_t$ whose value is $1$ if $X_t < \alpha X_{t-1} \ \wedge \ X_{t-1} < M$ and $0$ otherwise. Let us say that a time step $t$ is \emph{bad} if $Y_t = 1$ and \emph{good} if $Y_t = 0$. By hypothesis, we have
$$
\Prob{Y_t = 1 \;|\; X_{t-1} < M, X_{t-2}, \dots, X_1} \leqslant p
$$
Consider the first $t$ time steps and observe that if there are $\tau$ bad time steps and $t-\tau$ good time steps, then either $X_t \geqslant \alpha^{t-\tau} \beta^\tau$ or a time step $i \leqslant t$ exists such that $X_i \geqslant M$. So if $X_i < M$ for every $i = 1, \ldots, t$, then the number of bad time steps must be at least $\tau_t$ where
$$
\tau_t = \frac{t \log \alpha - \log M}{\log (\alpha/\beta)}
$$
Hence, if $t\geqslant \frac{\log M}{\log \alpha - e \, p \log(\alpha / \beta)}$  it holds that
$$
\Prob{\bigcap_{i=1}^t \{X_i < M\}} \leqslant \Prob{\sum_{i=1}^t Y_i \geqslant \tau_t} \leqslant \Prob{B(t,p) \geqslant \tau_t} \leqslant \Expf{-pt}
$$
where in the second inequality we used Lemma~\ref{lemma:ba}, in the last inequality Observation~\ref{obs:scb}, and the fact that the hypothesis on $p$ implies that $\tau_t \geqslant e \, p \, t$.
\qed

\noindent The following lemmas allow us to apply the previous lemma to the flooding process.

\noindent We say that a cell $C$ is \emph{infected} at time $t$ if, immediately after the move action of time step $t$, $C$ contains at least one informed node. The next lemma will be used in the Filling phase as well.

\begin{lemma}\label{infected-cells}
Positive constants $a$ and $b$ exist such that, for any time step $t$ and for any supercell $S$, if at the beginning of time step $t$ a supercell $S' \in N(S)$ contains $m$ informed nodes, then
\[
\Prob{Z \leqslant a m'} \; \leqslant \; \Expf{-b m'}
\]
where $Z$ is the random variable  counting the number of infected cells of $S$ at time $t$ and $m' = \min \{ m, L^2 / \ell^2 \}$.
\end{lemma}
\proof Let $S'$ be a supercell in $N(S)$, let $I$ be the set of informed nodes in $S'$, and set $m = |I|$. For any cell $C$ of $S$, let $Z_C$ be the $0-1$ random variable that has value $1$ iff immediately after the move action of time step $t$ at least one of the nodes in $I$ is in $C$. Let $Z' = \sum_C Z_C$. Clearly, it holds that $Z \geqslant Z'$. Since the $Z_C$'s are not independent, we will use the method of bounded differences (Lemma~\ref{mobd-average}) in order to prove that $Z'$ is concentrated around its expected value. Firstly, observe that
\[
\Prob{ Z_C = 0 } \; \leqslant \; \left(1 - \frac{\ell^2}{\pi \rho^2}\right)^m.
\]
For the sake of convenience let $\lambda = \left(1 - \frac{\ell^2}{\pi \rho^2}\right)$. It holds that
\[
\Expec{Z'} \;=\; \sum_C \Prob{Z_C = 1} \;\geqslant\; \frac{L^2}{\ell^2}(1 - \lambda^m).
\]
Let indicate the nodes in $I$ by the integers $1,2,\ldots, m$. For any $i\in I$, let $X_i$ be the random variable whose value is the position of node $i$ immediately after the move action (at time step $t$). Let $f : \prod_{i = 1}^m L_{n, \epsilon} \rightarrow \mathbb{R}$ be the function such that $f(\mathbf{x}_1,\ldots,\mathbf{x}_m)$ is the number of cells $C$ of $S$ such that, for some $i$, position $\mathbf{x}_i$ is in $C$. Clearly, $Z'$ is equal to the random variable $f(X_1,\ldots, X_m)$. Consider any integer $k$ with $1 \leqslant k \leqslant m$ and any positions $\mathbf{x}_1, \ldots, \mathbf{x}_{k-1}, \mathbf{x}_k, \mathbf{x}_k' \in L_{n, \epsilon}$. Let $\hat{X}_1,\ldots,\hat{X}_m$ be the random variables such that for $i = 1,\ldots, k$ $\hat{X}_i = \mathbf{x}_i$ and for $j = k+1,\ldots, m$ $\hat{X}_j = X_j$. Moreover, define random variables $Y$ and $Y'$ as follows
\[
Y\; =\; f(\hat{X}_1, \ldots, \hat{X}_m)\quad\mbox{and} \quad Y' \;=\; f(\hat{X}_1,\ldots,\hat{X}_{k-1},  \mathbf{x}_k', \hat{X}_{k+1},\ldots, \hat{X}_m).
\]
It is immediate to see that
\[
\begin{array}{l}
\Expec{Y} \;=\; \Expec{f(X_1\ldots, X_m) \;|\; (X_1,\ldots,X_{k-1})= (\mathbf{x}_1,\ldots, \mathbf{x}_{k-1}), X_k = \mathbf{x}_k} \\
\Expec{Y'} \;=\; \Expec{f(X_1\ldots, X_m) \;|\; (X_1,\ldots,X_{k-1})= (\mathbf{x}_1,\ldots, \mathbf{x}_{k-1}), X_k = \mathbf{x}_k'}
\end{array}
\]
Thus, we have to bound $|\Expec{Y} - \Expec{Y'}|$. Let $\zeta$ and $\zeta'$ be the number of cells of $S$ that contain some position in $\{\mathbf{x}_1,\ldots, \mathbf{x}_{k-1}, \mathbf{x}_k\}$ and $\{\mathbf{x}_1,\ldots, \mathbf{x}_{k-1}, \mathbf{x}_k' \}$, respectively. It is easy to see that if $\zeta = \zeta'$ then $\Expec{Y} = \Expec{Y'}$ and so $|\Expec{Y} - \Expec{Y'}| = 0$. If, instead, $\zeta \neq \zeta'$ then it must be the case that $|\zeta - \zeta'| = 1$. Without loss of generality, suppose that $\zeta = \zeta' + 1$. Consider now the analogue of variables $Z_C$
for the variables $\hat{X}_1,\ldots,\hat{X}_m$ and $\hat{X}_1,\ldots,\hat{X}_{k-1}, \mathbf{x}_k', \hat{X}_{k+1},\ldots, \hat{X}_m$. Call them $\hat{Z}_C$ and $\hat{Z}_C'$, respectively. Clearly, $\Expec{Y} = \sum_C \hat{Z}_C$ and $\Expec{Y'} = \sum_C \hat{Z}_C'$. Since $\zeta = \zeta' + 1$, there exists a cell $\hat{C}$ of $S$ such that $\hat{C}$ does not contain any position in $\{\mathbf{x}_1,\ldots, \mathbf{x}_{k-1}, \mathbf{x}_k' \}$, it contains $\mathbf{x}_k$, and for every other cell $C\neq \hat{C}$ of $S$ $\hat{Z}_C = \hat{Z}_C'$. This implies that
\[
\Expec{Y} - \Expec{Y'} \;=\; \Prob{\hat{Z}_{\hat{C}} = 1} - \Prob{\hat{Z}_{\hat{C}}' = 1} \; = \; 1 - \Prob{\hat{Z}_{\hat{C}}' = 1}.
\]
Moreover, it holds that
\[
1 - \Prob{\hat{Z}_{\hat{C}}' = 1} \; =\; \Prob{\hat{Z}_{\hat{C}}' = 0} \;\leqslant\; \lambda^{m - k}.
\]
We have thus proved that, in any case, it holds that
\[
\begin{array}{l}
| \Expec{f(X_1\ldots, X_m) \;|\; (X_1,\ldots,X_{k-1})= (\mathbf{x}_1,\ldots, \mathbf{x}_{k-1}), X_k = \mathbf{x}_k} \\
\mbox{ } - \Expec{f(X_1\ldots, X_m) \;|\; (X_1,\ldots,X_{k-1})= (\mathbf{x}_1,\ldots, \mathbf{x}_{k-1}), X_k = \mathbf{x}_k'} | \;\leqslant\; \lambda^{m - k}
\end{array}
\]
By applying Lemma~\ref{mobd-average} with $\delta = \frac{\Expec{Z'}}{2}$, we obtain that
\begin{small}
\[
\Prob{Z' \leqslant \frac{\Expec{Z'}}{2}} \;\leqslant\; \Prob{| Z' - \Expec{Z'}| \geqslant \frac{\Expec{Z'}}{2}} \;\leqslant\;
2\Expf{- \frac{\Expec{Z'}^2}{2\sum_{k = 1}^m (\lambda^{m - k})^2}}.
\]
\end{small}
Since $\Expec{Z'} \geqslant \frac{L^2}{\ell^2}(1 - \lambda^m)$, by simple calcutations, we get
\[
\frac{\Expec{Z'}^2}{2\sum_{k = 1}^m (\lambda^{m - k})^2} \geqslant \frac{\rho^2(1 - \lambda^m)}{4072\cdot\ell^2}
\]
Moreover, since $m \geqslant m'$ and $m' \leqslant \frac{L^2}{\ell^2}$, it holds that
\[
(1 - \lambda^m) \geqslant \frac{\ell^2}{2\pi \rho^2}m'
\]
This implies that
\[
\frac{\Expec{Z'}^2}{2\sum_{k = 1}^m (\lambda^{m - k})^2} \geqslant \frac{m'}{25586} \quad \mbox{ and } \quad \frac{\Expec{Z'}^2}{2} \geqslant   \frac{m'}{227}
\]
Finally,  we have that
\[
\Prob{Z \leqslant a m'} \leqslant \Prob{Z \leqslant a m'} \leqslant  \Expf{-b m'}  \] where $a = \frac{1}{227}$ and $b = \frac{1}{25586}$.
\qed

\noindent We can now fix the constant $\gamma$ defining quasi-informed supercells: $\gamma = \frac {a \eta}{227}$. For any supercell $S$, let $m_t(S)$ be the number of informed in $S$ at time step $t$.

\noindent For any time step $t$ let $Y_t = \max \{m_t(S)\, :\, S \mbox{ is a supercell} \}$.

\begin{lemma}\label{lemma:first}
For any time step $t$, it holds that
$$
\Prob{Y_{t+1} < 2 Y_t \;|\;  Y_t < \frac{r_0^2}{28}} \leqslant \Expf{-\frac{r_0^2}{224}}
$$
\end{lemma}
\proof Consider an informed node, and let $C$ be the cell containing that node immediately after the move action at time $t$. The expected number of nodes in $C$ is $\Expec{X} = \ell^2 \geqslant \frac{r_0^2}{28}$ and by using Chernoff bounds it holds that
$$
\Prob{X \leqslant \frac{r_0^2}{56}} \leqslant \Expf{- \frac{r_0^2}{224}}
$$
Clearly all the nodes in cell $C$ will be informed after the transmission action, hence
\begin{eqnarray*}
\Prob{Y_{t+1} \leqslant 2 Y_t \;|\; Y_t < \frac{r_0^2}{2 \cdot 56}} & \leqslant & \Prob{Y_{t+1} \leqslant \frac{r_0^2}{56}} \\
& \leqslant & \Prob{X \leqslant \frac{r_0^2}{56}} \leqslant \Expf{-\frac{r_0^2}{224}}
\end{eqnarray*}
\qed

\begin{lemma}\label{lemma:second}
For any supercell $S$ and for any time step $t$, it holds that
\[
\Probc{m_{t+1}(S) < 2 m_t(S)}{\frac{r_0^2}{28} < m_t(S) < \gamma \rho^2} \leqslant 3 \Expf{- \frac{b r_0^2}{56}}
\]
\end{lemma}
\proof For sake of convenience, define $\mathcal{H}_t$ to be the event
\[
\mathcal{H}_t = \left\{ \frac{r_0^2}{28} < m_t(S) < \gamma \rho^2 \right\}
\]
For every cell $C$ in supercell $S$ let $Z_C = 1$ if at least one of the $m_t(S)$ informed nodes will be in cell $C$ immediately after the move action of time $t$ and $Z_C = 0$ otherwise. Let $Z = \sum_{C \in S} Z_C$ be the number of cells in $S$ \emph{infected} by the $m_t(S)$ nodes. And let
\[
A = \ell^2 Z \geqslant \frac{r_0^2}{(1+\sqrt{2})^2} Z
\]
be the size of the \emph{infected area} in $S$ after the move action. \\
Now consider a supercell $S'$ adjacent to $S$. For every node $u$ in $S'$, at the beginning of time $t$, let $X_u = 1$ if $u$ will be in the infected area immediately after the move action and $X_u = 0$ otherwise, and let $X = \sum_{u \in S'} X_u$. Observe that all such nodes will be informed after the transmission action, hence $m_{t+1}(S) \geqslant X$. In what follows we will evaluate the probability that $X$ is less than $2$ times $m_t(S)$. Since, by assumption, event $\sD$ holds, we need to evaluate the following
\[
\Probc{X < 2 m_t(S)}{\mathcal{H}_t, \, \mathcal{D}}
\]
We first  write down the above probability in a more suitable way
\[
\Probc{X < 2 m_t(S)}{\mathcal{H}_t, \, \mathcal{D}}
=
\]
\[
= \ \sum_{\frac{r_0^2}{28} < m < \gamma \rho^2} \Probc{X < 2 m}{m_t(S) = m, \, \mathcal{D}} \Probc{m_t(S) = m}{\mathcal{H}_t, \, \mathcal{D}}
\]
If we prove that for every $m$ between $r_0^2/28$ and $\gamma \rho^2$ it holds that
\[
\Probc{X < 2 m}{m_t(S) = m, \, \mathcal{D}} \leqslant 3 \Expf{-\frac{r_0^2}{56}}
\]
then the thesis follows. Let us then relate the above probability to the size of the infected area\footnote{Here and in the sequel we will repeatedly use that for three events $\mathcal{A}, \mathcal{B}, \mbox{ and } \mathcal{C}$ we can write $\Probc{\mathcal{A}}{\mathcal{C}} \leqslant   \Probc{\mathcal{A}}{\mathcal{B}, \, \mathcal{C}} + \Probc{\overline{\mathcal{B}}}{\mathcal{C}}$.}
\begin{eqnarray}\label{eq:splitX}
\Probc{X < 2 m}{m_t(S) = m, \, \mathcal{D}} & \leqslant & \Probc{X < 2 m}{m_t(S) = m, \, \mathcal{D},\, A \geqslant \alpha m} + \nonumber \\
& & + \; \Probc{A < \alpha m}{m_t(S) = m, \mathcal{D}}
\end{eqnarray}
where $\alpha$ is a suitable constant that we will choose later. Now we evaluate the first term of the above sum. For every node $u$ in $S'$ it holds that
\begin{equation}\label{eq:eachnode}
\Prob{X_u = 1 \;\left|\; m_t(S) = m, \, \mathcal{D},\, A  \geqslant \alpha m \right. } \geqslant \frac{\alpha }{\pi \rho^2} m
\end{equation}
Hence the conditional expectation of $X$ is
\begin{equation}\label{eq:allexpec}
\Expec{X \;\left|\; m_t(S) = m, \, \mathcal{D},\, A \geqslant \alpha m \right. } \geqslant  \sum_u \frac{\alpha }{\pi \rho^2} m \geqslant \eta \rho^2 \frac{\alpha }{\pi \rho^2} m = \frac{\alpha \eta}{\pi} m
\end{equation}
If the $X_u$s were independent, by using Chernoff bound we could get
\[
\Prob{\left. X < \frac{\alpha \eta}{2 \pi} m \;\right|\; m_t(S) = m, \, \mathcal{D},\, A \geqslant \alpha m} \leqslant \Expf{- \frac{\alpha \eta}{8 \pi} m}
\]
Unfortunately, since event $\mathcal{D}$ makes assumption \emph{on the future}, r.v. $X_u$s conditioned on $\mathcal{D}$ are not independent anymore. Anyway, this is easy to handle. Split the event $\mathcal{D} = \mathcal{D}_1 \cap \mathcal{D}_2$ where $\mathcal{D}_1$ is the event ``The density condition holds since time step $t$'' and $\mathcal{D}_2$ is the event ``The density condition holds from time step $t+1$ to time step $n$''. Since $\mathcal{D}$ holds with high probability, then events $\mathcal{D}_1 \setminus \mathcal{D}$ and $\mathcal{D}_2 \setminus \mathcal{D}$ have negligible probabilities, hence conditioning on $\mathcal{D}$ is \emph{almost the same} as conditioning on $\mathcal{D}_1$. Formally, we can observe that for every event $\mathcal{A}$ it holds that $\Probc{\mathcal{A}}{\mathcal{D}} \leqslant 2 \Probc{\mathcal{A}}{\mathcal{D}_1}$.

\noindent By conditioning on $\mathcal{D}_1$ instead of $\mathcal{D}$, bounds~(\ref{eq:eachnode}) and~(\ref{eq:allexpec}) remains unchanged, and now we can apply Chernoff bound because the $X_u$s conditioned on $\mathcal{D}_1$ are independent
\[
\Prob{\left. X < \frac{\alpha \eta}{2 \pi} m \;\right|\; m_t(S) = m, \, \mathcal{D}_1,\, A  \geqslant \alpha m} \leqslant \Expf{- \frac{\alpha \eta}{8 \pi} m} \leqslant \Expf{- \frac{\alpha \eta}{224 \pi} r_0^2}
\]
And
\[
\Prob{\left. X < \frac{\alpha \eta}{2 \pi} m \;\right|\; m_t(S) = m, \, \mathcal{D},\, A  \geqslant \alpha m} \leqslant 2 \Expf{- \frac{\alpha \eta}{224 \pi} r_0^2}
\]
By choosing $\alpha = \frac{4 \pi}{\eta}$, we have that
\begin{equation}\label{eq:Xcond}
\Prob{\left. X < 2 m \;\right|\; m_t(S) = m, \, \mathcal{D},\, A  \geqslant \alpha m} \leqslant 2 \Expf{- \frac{r_0^2}{56}}
\end{equation}

\noindent As for the second addend of~(\ref{eq:splitX}), i.e. $\Probc{A < \alpha m}{m_t(S) = m, \mathcal{D}}$, two cases may arise.

\smallskip
\noindent \underline{Case 1:} $r_0^2/28 < m \leqslant L^2/\ell^2$.

\noindent From Lemma~\ref{infected-cells}, it holds that
\[
\Probc{Z \leqslant a m}{m_t(S) = m, \mathcal{D}} \;\leqslant\; \Expf{-b m} \leqslant \Expf{-b \frac{r_0^2}{28}}
\]
where $a$ and $b$ are the constants fixed by Lemma~\ref{infected-cells}. Hence
\begin{eqnarray*}
\Probc{A < \alpha m}{m_t(S) = m, \mathcal{D}} & = & \Probc{Z < \frac{\alpha}{\ell^2}m}{m_t(S) = m, \mathcal{D}} \\
& \leqslant & \Probc{Z < \frac{\alpha (1+\sqrt{2})^2}{r_0^2}m}{m_t(S) = m, \mathcal{D}}
\end{eqnarray*}

\noindent Thus, for $r_0 \geqslant (1+\sqrt{2}) \sqrt{\alpha/a}$, it holds that
\begin{small}
\[
\Probc{Z < \frac{\alpha (1+\sqrt{2})^2}{r_0^2}m}{m_t(S) = m, \mathcal{D}} \leqslant \Probc{Z \leqslant a m}{m_t(S) = m, \mathcal{D}} \leqslant \Expf{-b \frac{r_0^2}{28}}
\]
\end{small}

\medskip
\noindent \underline{Case 2:} $L^2/\ell^2 < m < \gamma \rho^2$.

\noindent From Lemma~\ref{infected-cells}, it holds that
\[
\Probc{Z \leqslant a L^2 / \ell^2}{m_t(S) = m, \mathcal{D}} \;\leqslant\; \Expf{-b \frac{L^2}{\ell^2}}
\]
where $a$ and $b$ are the constants fixed by Lemma~\ref{infected-cells}. Hence
\begin{eqnarray*}
\Probc{A < \alpha m}{m_t(S) = m, \mathcal{D}} & \leqslant & \Probc{A < \alpha \gamma \rho^2}{m_t(S) = m, \mathcal{D}} \\
& \leqslant & \Probc{A < 18 \alpha \gamma L^2}{m_t(S) = m, \mathcal{D}} \\
& = & \Probc{Z < 18 \alpha \gamma \frac{L^2}{\ell^2}}{m_t(S) = m, \mathcal{D}}
\end{eqnarray*}
Since $\gamma \leqslant \frac{a}{18 \alpha}$, it holds that
\[
\Probc{Z < 18 \alpha \gamma \frac{L^2}{\ell^2}}{m_t(S) = m, \mathcal{D}} \leqslant \Expf{-b \frac{L^2}{\ell^2}}
\]

\medskip
\noindent In both cases, since $L^2 /\ell^2 \geqslant r_0^2 / 28$, we get
\begin{equation}\label{eq:areacond}
\Probc{A < \alpha m}{m_t(S) = m, \mathcal{D}} \leqslant \Expf{-b \frac{r_0^2}{28}}
\end{equation}

\noindent By using~(\ref{eq:Xcond}) and~(\ref{eq:areacond}) in~(\ref{eq:splitX}), we obtain
\[
\Probc{X < 2 m}{m_t(S) = m, \, \mathcal{D}} \leqslant 3 \Expf{-b\frac{r_0^2}{56}}
\]
for every $r_0^2 < m < \gamma \rho^2$.
\qed

\begin{lemma} \label{lm::bootstrap}
Within $\mathcal{O}(\log n)$ time steps there is a quasi-informed supercell w.h.p.
\end{lemma}
\proof By using simple geometric arguments, if we choose $\beta = 1/121$, then it holds that $Y_{t+1} \geqslant \beta Y_t$ with probability one, thus satisfying Hypothesis (2) of Lemma~\ref{lemma:almostincreasing}. From Lemmas~\ref{lemma:first} and~\ref{lemma:second}, it is easy to see that the r.v. $Y_t$s satisfy Hypothesis (1) of Lemma~\ref{lemma:almostincreasing} with constants $\alpha = 2$, $M = \gamma \rho^2$ and $p = 3 \Expf{-b\frac{r_0^2}{224}}$. Hence, for a sufficiently large constant $r_0$, the thesis follows by applying Lemma~\ref{lemma:almostincreasing}.
\qed

\subsection{The spreading phase}
\begin{lemma}[Local-supercell spreading]\label{lemma:stability}
For any supercell $S$, if $m_t(S) \geqslant \gamma \rho^2$ then the event $m_{t+1}(S') \geqslant \gamma \rho^2$  holds for any supercell $S'\in N(S)$ with probability at least $1 - 1/ n ^4$.
\end{lemma}
\proof Let $S'$ be any supercell in $N(S)$. Since $m_t(S) \geqslant \gamma \rho^2 \geqslant \rho^2 /r_0^2$ (remind that $r_0$ is a constant that we can set sufficiently large), from Lemma~\ref{infected-cells}, it follows that
\[
\Prob{Z \leqslant a \frac{\rho^2}{ r_0^2}} \;\leqslant\; \Expf{-b\frac{\rho^2}{ r_0^2}}
\]
where $Z$ is the random variable counting the number of infected cells of $S'$ at time $t$.\\
Let $A = \ell^2 Z \geqslant \frac{r_0^2}{7}Z$ be the size of the \emph{infected area}. From the above inequality, it holds that

\begin{equation}\label{eq::infectedarea}
\Prob{A \leqslant a \frac{\rho^2}{7}} \;\leqslant\; \Expf{-b \frac{\rho^2}{  r_0^2}}
\end{equation}

\noindent Now consider a super-cell $S''$ adjacent to $S'$ and different from $S$. We want to evaluate how many nodes from $S''$ move to the infected area. For every node $u$ in $S''$ let $X_u$ be the binary r.v. whose value is $1$ iff node $u$, after the move action, is in the infected area of $S'$. Since we are assuming the density event $\sD$ holds, the r.v. $X_u$'s are not independent. We thus first bound the relative probabilities under the weaker assumption that the density condition holds till time step $t-1$: This makes r.v. $X_u$'s independent. Then, we bring this bound under the density condition $\sD$.

\noindent It is easy to see that
\[
\Probc{X_u =1}{A \geqslant a  \frac{\rho^2}{7 }} \geqslant a \frac 1{7 \pi}
\]
Since the density condition holds before the move action of time $t$,  we get

\[
\Expec{X \left|\; A \geqslant a \frac{\rho^2}{7 } \right.} \geqslant a \frac{\eta \rho^2}{7 \pi}
\]
where $X$ is the r.v. counting the number of nodes of $S''$ moving to the infected area of $S'$. Observe that $X$ is a lower bound on the number of informed node in cell $S''$ after the transmission action of time step $t$. By applying Chernoff's bound, we obtain
\[
\Probc{X \leqslant a \frac{\eta \rho^2}{14\pi}}{A \geqslant a \frac{\rho^2}{7 }} \leqslant \Expf{- a \frac{\eta\rho^2}{56 \pi}}
\]
Then, from (\ref{eq::infectedarea}), we get
\begin{eqnarray*}
\Prob{X \leqslant a \frac{\eta \rho^2}{14 \pi}} & \leqslant & \Probc{X \leqslant a \frac{\eta \rho^2}{14\pi}}{A \geqslant a \frac{\rho^2}{7 }} +  \Prob{A \leqslant a \frac{\rho^2}{7 }} \\
& \leqslant & \Expf{- a \frac{\eta\rho^2}{56\pi}} + \Expf{- b \frac{\rho^2}{ r_0^2}} \leqslant \frac 1{n^5}
\end{eqnarray*}

\noindent for a suitable choice of constant $c$ such that $\rho \geqslant c \sqrt{\log n}$. Since $\gamma \leqslant a \frac{\eta }{14\pi}$, we obtain
\[
\Probc{X \geqslant \gamma  \rho^2}{\sD} \geqslant (1 - \frac 1{n^5} ) -  2 \Prob{\overline \sD} \geqslant 1 - \frac 1 {n^4}
\]
\qed

\noindent In order to prove a bound on the number of time steps that are sufficient to guarantee (with high probability) that one quasi-informed supercell spreads the information to all the supercells, we need two lemmas. The first one will provide a bound on the number of supercells that are adjacent to an arbitrary set of supercells.

\noindent Let $Q$ be a $m\times m$ square grid, that is, $Q$ is a square partitioned into $m\times m$ congruent sub-squares, called cells. For any subset $B$ of the cells of $Q$, define the \emph{boundary}  $\partial B$ of $B$ as the set of all the cells that do not belong to $B$ and that are adjacent to some cell in $B$:
\[
\partial B = \{ c \;|\; c \not\in B \;\wedge\; \exists c'\in B: \mbox{ $c'$ is adjacent to $c$}\}.
\]

\begin{lemma}[Boundary size]\label{boundary}
Let $Q$ be a $m \times m$ square grid and let $B$ be any subset of the cells of $Q$. It holds that
\[
|\partial B| \;\geqslant\; \sqrt{\min\{|B|, m^2 - |B|\}}.
\]
\end{lemma}
\proof For the sake of convenience, we say that a cell in $B$ is a black cell and all the cells not in $B$ are white cells. We say that a row of the grid $Q$ is black if all the cells of the row are black. Similarly, we define a black column. Moreover, a row or a column which contains both at least one black cell and at least one white cell is said to be gray. Let $b_r$ and $b_c$ be, respectively, the number of black rows and the number of black columns. To prove the bound on $|\partial B|$ we distinguish four cases.
\begin{description}
\item[$b_r = 0 \wedge b_c \geqslant 1$:] In this case, every row is gray. This implies that every row contains at least one cell in $\partial B$ (if the leftmost cell of the row is black then the leftmost white cell belongs to $\partial B$ otherwise the cell immediately to the left of the leftmost black cell belongs to $\partial B$). Since the rows are $m$, it follows that $|\partial B| \geqslant m$. It immediately derives that  $|\partial B| \;\geqslant\; \sqrt{\min\{|B|, m^2 - |B|\}}$, since $\sqrt{\min\{|B|, m^2 - |B|\}} \leqslant \sqrt{m^2} = m$.

\item[$b_r \geqslant 1 \wedge b_c = 0$:] This case is symmetric to the previous one.

\item[$b_r \geqslant 1 \wedge b_c \geqslant 1$:] Without loss of generality, assume that $b_r \leqslant b_c$. Since $b_c \geqslant 1$, there are $m - b_r$ gray rows. Thus, $|\partial B| \geqslant m - b_r$. It is easy to see that the number of cells belonging to either black rows or black columns is $m\cdot b_r + m\cdot b_c - b_r\cdot b_c$. Since these are black cells, it holds that
\[
|B|\; \geqslant\; m\cdot b_r + m\cdot b_c - b_r\cdot b_c \;=\; m\cdot b_r + b_c(m - b_r) \;\geqslant\; m\cdot b_r + b_r(m - b_r).
\]
Now, it easy to verify that the inequality $|B| \geqslant m\cdot b_r + b_r(m - b_r)$ implies that
\[
b_r \;\leqslant\; m - \sqrt{m^2 - |B|}.
\]
Hence, it holds that $|\partial B| \geqslant m - b_r \geqslant m - (m - \sqrt{m^2 - |B|}) = \sqrt{m^2 - |B|}$, and thus $|\partial B| \geqslant \sqrt{\min\{|B|, m^2 - |B|\}}$.

\item[$b_r = 0 \wedge b_c = 0$:] Let $y_r$ and $y_c$ be, respectively, the number of gray rows and the number of gray columns. Since there are neither black rows nor black columns, it must be the case that every black cell belongs to both a gray row and a gray column. This implies that
\[
y_r\cdot y_c \;\geqslant |B|.
\]
Without loss of generality, assume that $y_r \geqslant y_c$. It follows that $y_r^2 \geqslant |B|$, and so $y_r \geqslant \sqrt{|B|}$. Since every gray row contains at least a cell in $\partial B$, it holds that $|\partial B| \geqslant \sqrt{|B|} \geqslant \sqrt{\min\{|B|, m^2 - |B|\}}$.
\end{description}
\qed

The second lemma will allow us to prove an upper bound on the number of steps to get all the supercells quasi-informed when, in one time step, the information propagates from all the quasi-informed supercells to their adjacent ones.

\begin{lemma}[Spreading time I]\label{spreading-time}
Let $K$ be any integer with $K\geqslant 1$ and let $\{q_t \;|\; t\in \mathbb{N}\}$ be a succession of integers such that $q_0 \geqslant 1$, for every $t\geqslant 0$, $q_t \leqslant K$ and $q_{t+1} \geqslant q_t + \sqrt{\min\{q_t, K- q_t\}}$. Then, it holds that, for every $t \geqslant 5\sqrt{K}$, $q_t = K$.
\end{lemma}
\proof If $K \leqslant 2$ it is immediate to see that the thesis holds, so in the sequel of the proof we assume that $K \geqslant 3$. Firstly observe that if $q_t \leqslant K - 1$ then $q_{t+1} \geqslant q_t + 1$. This implies that an integer $\tau \geqslant 0$ exists such that $\tau = \min\{ t \;|\; q_t \geqslant K/2\}$. Now we prove, by induction on $t$, that, for every $t$ with $0 \leqslant t \leqslant \tau$, $q_t \geqslant \frac{t^2}{8}$. For $t = 0, 1, 2$ it is trivially true. Assume that it is true for $t \geqslant 2$ (and $t + 1\leqslant \tau$). Then, since $q_t < K/2$, it holds that
\[
q_{t+1} \;\geqslant\; q_t + \sqrt{q_t} \; \geqslant\; \frac{t^2}{8} + \sqrt{\frac{t^2}{8}} \;=\; \frac{t^2 + \sqrt{8}t}{8} \; \geqslant\; \frac{(t+1)^2}{8}
\]
and this completes the proof by induction. From this it follows that if $\tau \geqslant 1$ then $q_{\tau - 1} \geqslant \frac{(\tau - 1)^2}{8}$. Thus, it is easy to verify that $\tau \leqslant 2\sqrt{K} + 1$. Now, we prove, by induction on $t$, that
\[
\forall t \;:\; \tau \leqslant t \leqslant \left\lceil 5\sqrt{K}\right\rceil \qquad q_t \;\geqslant\; K - \left\lceil \frac{\left(\left\lceil 5\sqrt{K}\right\rceil - t\right)^2}{9}\right\rceil.
\]
For $t = \tau$ it holds that
\begin{eqnarray*}
K - \left\lceil \frac{\left(\left\lceil 5\sqrt{K}\right\rceil - \tau\right)^2}{9}\right\rceil &\leqslant &
K - \frac{\left(5\sqrt{K} - 2\sqrt{K} - 1\right)^2}{9} \quad \mbox{(since $\tau \leqslant 2\sqrt{K} + 1$)}\\
& = & \frac{9K - \left(3\sqrt{K} - 1\right)^2}{9} \; =\; \frac{2\sqrt{K}}{3} -\frac{1}{9} \;\leqslant\; \frac{K}{2} \;\leqslant\; q_{\tau}.
\end{eqnarray*}
Assume now that it is true for $t \geqslant \tau$ (and $t + 1\leqslant \lceil 5\sqrt{K}\rceil$). This implies that an integer $\delta \geqslant 0$ exists such that
\[
q_t \; = \; K - \left\lceil \frac{h^2}{9}\right\rceil + \delta
\]
where we set $h = \lceil 5\sqrt{K}\rceil - t$. Since $q_t \geqslant K/2$, it holds that
\[
q_{t+1} \;\geqslant\; q_t + \sqrt{K - q_t} \;=\; K - \left\lceil\frac{h^2}{9}\right\rceil + \delta + \sqrt{\left\lceil\frac{h^2}{9}\right\rceil  - \delta}.
\]
From the inequality $\forall a \geqslant b\geqslant 0$ $\sqrt{a - b} \geqslant \sqrt{a} - \sqrt{b}$, it derives that
\[
\delta +  \sqrt{\left\lceil\frac{h^2}{9}\right\rceil  - \delta} \;\geqslant \;  \delta +  \sqrt{\left\lceil\frac{h^2}{9}\right\rceil} - \sqrt{\delta} \;\geqslant\; \sqrt{\left\lceil\frac{h^2}{9}\right\rceil}
\]
where the last inequality (i.e. $\delta \geqslant \sqrt{\delta}$) holds since $\delta$ is an integer.
Thus,
\[
q_{t+1}\;\geqslant\; K - \left\lceil\frac{h^2}{9}\right\rceil  + \sqrt{\left\lceil\frac{h^2}{9}\right\rceil}.
\]
Now, we prove that, for every integer $j \geqslant 1$, the following inequality holds
\[
\sqrt{\left\lceil\frac{j^2}{9}\right\rceil} \;\geqslant\; \left\lceil\frac{j^2}{9}\right\rceil - \left\lceil\frac{(j - 1)^2}{9}\right\rceil.
\]
For $j = 1,2,\ldots, 7$, it can be exhaustively proved case by case. Let $j \geqslant 8$. It holds that
\[
\sqrt{\left\lceil\frac{j^2}{9}\right\rceil} \;\geqslant\; \sqrt{\frac{j^2}{9}} \;\geqslant\; \frac{j}{3} \;\geqslant\; \frac{8 + 2j}{9}
\]
where the last inequality is due to the assumption that $j \geqslant 8$. Moreover,
\[
\left\lceil\frac{j^2}{9}\right\rceil - \left\lceil\frac{(j - 1)^2}{9}\right\rceil \;\leqslant\; \frac{j^2}{9} + 1 - \frac{(j - 1)^2}{9} \; =\; \frac{8 + 2j}{9}.
\]
Thus the inequality is proved. By using it with $j = h$, we obtain that
\begin{small}
\[
q_{t+1}\;\geqslant\; K - \left\lceil\frac{h^2}{9}\right\rceil  + \sqrt{\left\lceil\frac{h^2}{9}\right\rceil} \;\geqslant\; K -  \left\lceil\frac{(h - 1)^2}{9}\right\rceil \;=\; K - \left\lceil \frac{\left(\left\lceil 5\sqrt{K}\right\rceil - (t + 1)\right)^2}{9}\right\rceil
\]
\end{small}
and this completes the proof by induction. Thus, we have proved that for $\hat{t} = \lceil5\sqrt{K} \rceil$, it holds that
\[
q_{\hat{t}} \;\geqslant\; K - \left\lceil \frac{\left(\left\lceil 5\sqrt{K}\right\rceil - \hat{t}\right)^2}{9}\right\rceil \; =\; K.
\]
Hence, for every $t \geqslant \lceil5\sqrt{K} \rceil$, it holds that $q_t = K$.
\qed

\noindent By combining Lemmas~\ref{lemma:stability}, \ref{boundary} and~\ref{spreading-time} we get the following bound.

\begin{lemma}[Spreading time II]\label{spreading}
If at time $t_1 \leqslant  n/2$, there is at least one quasi-informed supercell then, with probability at least $1 - \frac{1}{n^2}$, at every time $t$ with $t_1 + 22\frac{\sqrt{n}}{\rho} \leqslant t \leqslant n$ all the supercells are quasi-informed.
\end{lemma}
\proof In the sequel, when we write time $t$ we mean time $t_1 + t$. For any $t \geqslant 0$, let $\mathcal{Q}_t$ be the set of quasi-informed supercells at time $t$. By hypothesis $|\mathcal{Q}_0| \geqslant 1$. In virtue of Lemma~\ref{boundary}, if all the supercells in $\mathcal{Q}_t$ and all their adjacent supercells get quasi-informed, at time $t+1$, then
\[
|\mathcal{Q}_{t+1}| \geqslant |\mathcal{Q}_t| + \sqrt{\min\{ |\mathcal{Q}_t|, N -  |\mathcal{Q}_t|\} }
\]
where $N= \frac{n}{L^2}$ is the number of supercells. This implies that if the above inequality does not hold then there exists a supercell $S\in\mathcal{Q}_t$ such that either $S$ or one adjacent supercell of $S$ is not quasi-informed at time $t+1$. It follows that
\[
\Prob{|\mathcal{Q}_{t+1}| < |\mathcal{Q}_t| + \sqrt{\min\{ |\mathcal{Q}_t|, N -  |\mathcal{Q}_t|\} }}  \;\leqslant\; \Prob{\exists S\in \mathcal{Q}_t : \mathcal{E}_{S, t+1}}
\]
where $\mathcal{E}_{S, t+1}$ is the event that occurs if $S$ or one adjacent supercell of $S$ is not quasi-informed at time $t+1$. By the union bound, it holds that
\begin{small}
\[
\Prob{\exists S\in \mathcal{Q}_t : \mathcal{E}_{S, t+1}} \;\leqslant\; \sum_{S} \Prob{S\in \mathcal{Q}_t \wedge \mathcal{E}_{S, t+1}} \;=\; \sum_{S} \Probc{ \mathcal{E}_{S, t+1} }{ S\in \mathcal{Q}_t } \Prob{S\in \mathcal{Q}_t}.
\]
\end{small}

\noindent From Lemma~\ref{lemma:stability}, for every supercell $S$,
\[
\Probc{\mathcal{E}_{S, t+1} }{S\in \mathcal{Q}_t }\;\leqslant\; \frac{1}{n^4}.
\]
It follows that
\begin{eqnarray*}
\Prob{|\mathcal{Q}_{t+1}| < |\mathcal{Q}_t| + \sqrt{\min\{ |\mathcal{Q}_t|, N -  |\mathcal{Q}_t|\} }}  & \leqslant & \sum_{S} \Probc{\mathcal{E}_{S, t+1} }{S\in \mathcal{Q}_t } \Prob{S\in \mathcal{Q}_t} \\
& \leqslant &  \sum_{S} \Probc{\mathcal{E}_{S, t+1} }{ S\in \mathcal{Q}_t }\\
& \leqslant & \frac{N}{n^4} \;\leqslant\; \frac{1}{n^3}.
\end{eqnarray*}
Thus, by the union bound, with probability at least $1 - \frac{1}{n^2}$, it holds that
\[
\forall t = 0,1,\ldots, n/2\qquad |\mathcal{Q}_{t+1}| \;\geqslant\; |\mathcal{Q}_t| + \sqrt{\min\{ |\mathcal{Q}_t|, N -  |\mathcal{Q}_t|\} }.
\]
By applying Lemma~\ref{spreading-time} with $q_t = |\mathcal{Q}_t |$ and $K = N$, we obtain that, for every $t$ with $5\sqrt{N} \leqslant t \leqslant n/2$, $|\mathcal{Q}_t| = N$. The thesis follows by taking into account that $t_1 \leqslant n/2$, $t$ means time $t_1 + t$, and $5\sqrt{N} \leqslant 5\frac{\sqrt{n}}{3\sqrt{2}\rho} \leqslant 22\frac{\sqrt{n}}{\rho}$.
\qed

\subsection{The filling phase}
We first prove that a node not yet informed and belonging to a quasi-informed supercell will get informed in one time step, with a constant probability.

\begin{lemma}\label{singlenode}
There exists a constant $\beta > 0$ such that, for any node $u$, if at the beginning of a time step $t$ the supercell that contains $u$ is quasi-informed and node $u$ is not informed then, with probability at least $\beta$, node $u$ gets  informed by the end of time step $t$.
\end{lemma}
\proof Assume that at the beginning of time step $t$ node $u$ is not informed and the supercell $S$ that contains $u$ is quasi-informed. Thus $S$ contains at least $\gamma\rho^2 \geqslant \frac{\rho^2}{r_0^2}$ informed nodes at the beginning of time step $t$. From Lemma~\ref{infected-cells}, it derives that
\[
\Prob{Z \leqslant \frac{a\rho^2}{r_0^2}} \;\leqslant\; \Expf{-\frac{b\rho^2}{r_0^2}}
\]
where $Z$ is the random variable counting the number of infected cells of $S$ at time $t$. Now, let $\mathcal{U}$ be the event that occurs if node $u$ gets informed by the end of time step $t$. It is immediate to see that, for any $k \geqslant 0$,
\[
\Probc{ \mathcal{U}}{ Z = k} \;\geqslant\; \frac{k\ell^2}{\pi \rho^2}.
\]
It follows that
\begin{eqnarray*}
\Prob{\mathcal{U}} & = &\sum_{k \geqslant 0}  \Probc{ \mathcal{U}}{ Z = k}\Prob{Z = k} \\
& \geqslant & \sum_{k \geqslant \frac{a\rho^2}{r_0^2}} \Probc{ \mathcal{U}}{ Z = k}\Prob{Z = k}\\
& \geqslant & \frac{a}{6\pi}\sum_{k \geqslant \frac{a\rho^2}{r_0^2}} \Prob{Z = k} \\
& = & \frac{a}{6\pi}\Prob{Z \geqslant \frac{a\rho^2}{r_0^2}}\\
& \geqslant & \frac{a}{6\pi}\left(1 - \Expf{-\frac{b\rho^2}{r_0^2}}\right)\\
& \geqslant & \beta
\end{eqnarray*}
for a suitable constant $\beta > 0$.
\qed

\noindent From the above lemma we can derive a logarithmic upper bound for the filling time.

\begin{lemma}\label{filling}
If a time step $t_2 \leqslant \frac{3n}{4}$ exists such that at every time step $t$ with $t_2\leqslant t\leqslant n$ all the supercells are quasi-informed, then by time $t_2 + \mathcal{O}(\log n)$ all the nodes are informed, w.h.p.
\end{lemma}
\proof In the sequel, when we write time $t$ we mean time $t_2 + t$. Let $u$ be any node. For any time $t$, let $\mathcal{U}_t$ be the event that occurs if node $u$ is not informed by the end of time step $t$. Let $\beta$ be the constant of Lemma~\ref{singlenode} and let $k$ be a constant such that $\beta^{k\log n} < \frac{1}{n^2}$. By hypothesis, for every $t = 0, 1, \ldots k\log n$, all the supercells are quasi-informed. This implies that at each of these time steps the supercell that contains $u$ is quasi-informed. Thus, from Lemma~\ref{singlenode}, it holds that
\[
\Prob{\bigwedge_{t = 0}^{k\log n} \mathcal{U}_t} \; = \; \prod_{t = 0}^{k\log n} \Probc{\mathcal{U}_t}{ \bigwedge_{j = 0}^{t - 1} \mathcal{U}_j} \;\leqslant\; \beta^{k\log n} \; < \; \frac{1}{n^2}.
\]
This means that a node $u$ is not informed within $k\log n$ time steps with probability at most $\frac{1}{n^2}$. Hence, a direct application of the union bound shows that, with probability at least $1 - \frac{1}{n}$, all the nodes are informed within $k\log n$ steps.
\qed

\medskip
\noindent Finally, Theorem~\ref{thm:geoflooding} follows from Lemma~\ref{lm::bootstrap}, Lemma~\ref{spreading}, and Lemma~\ref{filling}.

\section{Conclusions}
Some interesting issues concerning the flooding time on geometric-MEG are still open. There is a logarithmic gap between our upper bound and the known lower bound~\cite{CMPS09} when the   move radius $\rho$ is very large. Closing this gap is an open problem.

\noindent A more challenging open issue is to extend our upper bound in the case where $\rho$ and $r$ are both very small (i.e. below $\sqrt{\log n}$).

\noindent Another open research work is to study the flooding time when the starting distribution is not the stationary one but is arbitrary (i.e. a worst-case analysis). We conjecture that the worst-case flooding time is not asymptotically larger than the stationary one for a large range of the network parameters.

\noindent Observe that our upper bound can be easily extended to the gossiping task (i.e. the all-to-all communication). It would be interesting to extend our analysis to other basic communication tasks such as data-gathering and routing.

\noindent Finally, a major challenge is to obtain results similar to our upper bounds for more realistic mobility models~\cite{MS09}.

\bibliographystyle{plain}
\bibliography{mrn}

\begin{thebibliography}{10}

\bibitem{AF99}
D.~Aldous and J.~Fill.
\newblock {\em Reversible Markov Chains and Random Walks on Graphs}.
\newblock http://stat-www.berkeley.edu/users/aldous/RWG/book.html, 2002.

\bibitem{A05}
C.~Amb{\"u}hl.
\newblock An optimal bound for the {MST} algorithm to compute energy efficient
  broadcast trees in wireless networks.
\newblock In {\em Proc. of 32th International Colloquium on Automata, Languages
  and Programming (ICALP)}, volume 3580 of {\em LNCS}, pages 1139--1150.
  Springer, 2005.

\bibitem{AKL08}
C.~Avin, M.~Koucky, and Z.~Lotker.
\newblock How to explore a fast-changing world.
\newblock In {\em Proc. of 35th International Colloquium on Automata, Languages
  and Programming (ICALP'08)}, volume 5125 of {\em LNCS}, pages 121--132.
  Springer, 2008.

\bibitem{ABKU99}
Y.~Azar, A.Z. Broder, A.R. Karlin, and E.~Upfal.
\newblock Balanced allocations.
\newblock {\em SIAM Journal on Computing}, 29(1):180--200, 1999.

\bibitem{CBD02}
T.~Camp, J.~Boleng, and V.~Davies.
\newblock A survey of mobility models for ad hoc network research.
\newblock {\em Wireless Communication and Mobile Computing}, 2(5):483--502,
  2002.

\bibitem{CKNR07}
I.~Chatzigiannakis, A.~Kinalis, S.~E. Nikoletseas, and J.~D.~P. Rolim.
\newblock Fast and energy efficient sensor data collection by multiple mobile
  sinks.
\newblock In {\em Proc. of MOBIWAC'07}, pages 25--32, 2007.

\bibitem{CMMPS08}
A.~Clementi, C.~Macci, A.~Monti, F.~Pasquale, and R.~Silvestri.
\newblock Flooding time in edge-markovian dynamic graphs.
\newblock In {\em Proc. of 27th Annual ACM SIGACT-SIGOPS Symposium on
  Principles of Distributed Computing (PODC'08)}, pages 213--222. ACM Press,
  2008.

\bibitem{CMPS09}
A.~Clementi, A.~Monti, F.~Pasquale, and R.~Silvestri.
\newblock Information spreading in stationary markovian evolving graphs.
\newblock In {\em Proc. of the 23rd IEEE International Parallel and Distributed
  Processing Symposium}. IEEE Computer Society, 2009.

\bibitem{DMP08}
J.~Diaz, D.~Mitsche, and X.~Perez-Gimenez.
\newblock On the connectivity of dynamic random geometric graphs.
\newblock In {\em Proc. of 19th annual ACM-SIAM symposium on Discrete
  algorithms (SODA'08)}, pages 601--610, 2008.

\bibitem{JetAl06}
S.~Jain et~Al.
\newblock Exploiting mobility for energy efficient data collection in wireless
  sensor networks.
\newblock {\em ACM/Kluwer Mobile Networks and Applications (MONET)}, 11(3),
  2006.

\bibitem{GT02}
M.~Grossglauser and N.C. Tse.
\newblock Mobility increases the capacity of ad-hoc wireless networks.
\newblock {\em IEEE/ACM Trans. on Networking}, 10(4), 2002.

\bibitem{G87}
R.A. Guerin.
\newblock Channel occupancy time distribution in a cellular radio system.
\newblock {\em IEEE Trans. on Veichular Technology}, 36(3):89--99, 1987.

\bibitem{GK99}
P.~Gupta and P.R. Kumar.
\newblock Critical power for asymptotic connectivity in wireless networks.
\newblock {\em Stochastic Analysis, Control, Optimization and Applications},
  pages 547--566, 1998.

\bibitem{KN08}
A.~Kinalis and S.~E. Nikoletseas.
\newblock Adaptive redundancy for data propagation exploiting dynamic sensory
  mobility.
\newblock In {\em Proc. of ACM MSWIM'08}, pages 149--156, 2008.

\bibitem{KKKP00}
L.~GC02M. Kirousis, E.~Kranakis, D.~Krizanc, and A.~Pelc.
\newblock Power consumption in packet radio networks.
\newblock {\em Theoretical Computer Science}, 243:289--305, 2000.

\bibitem{M89}
C.~McDiarmid.
\newblock On the method of bounded differences.
\newblock In {\em (J. Siemons ed.), London Mathematical Society Lecture Note},
  141, pages 148--188. Cambridge University Press, 1989.

\bibitem{MS09}
A.~Mei and J.~Stefa.
\newblock Swim: a simple model to generate small mobile worlds.
\newblock In {\em Proc. of IEEE INFOCOM'09}, 2009.

\bibitem{PPC06}
L.~Pelusi, A.~Passarella, and M.~Conti.
\newblock Beyond manets: Dissertation on opportunistic networking.
\newblock {\em IIT-CNR Tech. Rep.}, 2006.

\bibitem{Penrose}
M.~Penrose.
\newblock {\em Random Geometric Graphs}.
\newblock Oxford University Press, 2003.

\bibitem{SB03}
P.~Santi and D.~M. Blough.
\newblock The critical transmitting range for connectivity in sparse wireless
  ad hoc networks.
\newblock {\em IEEE Transactions on Mobile Computing}, 2(1):25--39, 2003.

\bibitem{Z06}
Z.~Zhang.
\newblock Routing in intermittently connected mobile ad-hoc networks and delay
  tolerant networks: overview and challenges.
\newblock {\em IEEE Communication Surveys}, 8(1), 2006.

\bibitem{ZAZ04}
W.~Zhao, M.~Ammar, and E.~Zegura.
\newblock A message ferrying approach for data delivery in sparse mobile ad-hoc
  networks.
\newblock In {\em Proc. of 5th ACM MobiHoc'04}, 2004.

\end{thebibliography}

\appendix
\section{Basic Probability}

\begin{lemma}[Chernoff's bound]\label{lemma:cb}
Let be $X = \sum_{i=1}^n X_i$ where $X_1, \dots, X_n$ are independent Bernoulli random variables and let be $0 < \varepsilon<1$. If  $0 < \mu \leqslant \mathbf{E}[X]$, then it
holds
\[
\mathbf{P} \{ X \leqslant (1-\varepsilon) \mu \} \leqslant e^{-\frac{\varepsilon^2}{2}\mu}
\]
\end{lemma}

\begin{obs}\label{obs:scb}
Let $X$ be a binomial random variable with $\Expec{X}=\mu$. Then
$$
\Prob{X \geqslant e \, \mu} \leqslant e^{-\mu}
$$
\end{obs}

\noindent We use the following standard probability bound (See~\cite{ABKU99}).
\begin{lemma}\label{lemma:ba}
Let $X_1, \dots, X_n$ be a sequence of random variables with values in an arbitrary domain, and let $Y_1, \dots, Y_n$ be a sequence of binary random variables, with the property that $Y_i = Y_i(X_1, \dots, X_i)$. If
$$
\Prob{Y_i = 1 \;|\; X_1, \dots, X_{i-1}} \leqslant p
$$
then
$$
\Prob{\sum Y_i \geqslant k} \leqslant \Prob{B(n,p) \geqslant k}
$$
where $B(n,p)$ denotes the binomially distributed random variable with parameters $n$ and $p$.
\end{lemma}

\noindent The following lemma states one inequality of the method of bounded differences (see Corollary~6.10 in~\cite{M89}).

\begin{lemma}\label{mobd-average}
Let $X_1,\ldots, X_n$ be random variables, with $X_k$ taking values in a set $A_k$ for each $k$, and let $\underline{X}$ denote the vector $(X_1,\ldots, X_n)$. Let $f : \prod_{k = 1}^n A_k \rightarrow \mathbb{R}$ be an appropriately measurable function. Suppose that there are constants $c_1,\ldots, c_n$ so that
\[
\begin{array}{l}
| \Expec{f(\underline{X}) \;|\; (X_1,\ldots,X_{k-1})= (x_1,\ldots, x_{k-1}), X_k = x_k} \\
\mbox{ } - \Expec{f(\underline{X}) \;|\; (X_1,\ldots,X_{k-1})= (x_1,\ldots, x_{k-1}), X_k = x_k'} | \;\leqslant\; c_k
\end{array}
\]
for each $k = 1,\ldots, n$ and $x_i\in A_i$ ($i = 1,\ldots, k-1$) and $x_k, x_k' \in A_k$. Then for any $\delta > 0$,
\[
\Prob{| f(\underline{X}) - \Expec{f(\underline{X})}| \geqslant \delta } \;\leqslant\; 2\Expf{-\frac{2\delta^2}{\sum_{k=1}^n c_k^2}}.
\]
\end{lemma}

\end{document}